\documentstyle[12pt]{article}

\newtheorem{theorem}{THEOREM}
\newtheorem{proposition}[theorem]{PROPOSITION}
\newtheorem{example}{EXAMPLE}
 
\newcommand{\spc}[1]{\mskip#1\thinmuskip}
\newcommand{\Sect}{Sect.}
\newcommand{\BAL}{{\it BAL}}
\newcommand{\ECL}{{\it ECL}}
\newcommand{\cplx}{{\bf C}}
\newcommand{\real}{{\bf R}}
\newcommand{\rint}{{\bf Z}}
\newcommand{\id}{{\rm id}}
\newcommand{\intdif}[1]{\partial^{#1}}
\newcommand{\pdmod}{\intdif{\rint}}
\newcommand{\ALGEB}{A}
\newcommand{\FUNCT}{F}
\newcommand{\IDEAL}{I}
\newcommand{\algea}{\overline{\cal A}_0}
\newcommand{\algeb}{\overline{\cal A}}
\newcommand{\funct}{\overline{\cal F}}
\newcommand{\vectf}{\overline{\cal X}}
\newcommand{\ideal}{\overline{\cal I}}
\newcommand{\idealvf}{
 \ideal\spc{.8}\overline{\makebox[.7em][r]{$\cal X$}}}
\newcommand{\alge}{{\cal A}}
\newcommand{\func}{{\cal F}}
\newcommand{\vect}{{\cal X}}
\newcommand{\mfdM}{{\it M}}
\newcommand{\Riem}[2]{\langle#1,\:#2\rangle}
\newcommand{\curvM}{c}
\newcommand{\Ders}{\frac{\partial}{\partial s}}
\newcommand{\nablaof}[1]{\nabla\!{}_{\!#1}\,}
\newcommand{\nablas}{\nablaof{s}}
\newcommand{\extd}{{\it d}}
\newcommand{\laT}{{\sf T}}
\newcommand{\laN}{{\sf N}}
\newcommand{\laB}{{\sf B}}
\newcommand{\laX}{{\sf X}}
\newcommand{\laY}{{\sf Y}}

\newcommand{\liN}{N}
\newcommand{\liB}{B}
\newcommand{\liX}{X}
\newcommand{\liY}{Y}
\newcommand{\liZ}{Z}
\newcommand{\embed}{\wp}
\newcommand{\canhom}{\varphi}
\newcommand{\forms}[1]{{\cal D}^{#1}}
\newcommand{\cycle}{{\rm cycle}}
\newcommand{\length}{\ell}
\newcommand{\dual}{{}^\ast}
\newcommand{\cMll}{\curvM\:\!\length^2}
 
\begin{document}
 
\def\thefootnote{\fnsymbol{footnote}}
 
\begin{flushright}
Preprint OCU-157, hep-th/9510172
\end{flushright}
\bigskip
\begin{center}
 \LARGE    Differential Geometry and Integrability of the\\
           Hamiltonian System of a Closed Vortex Filament\footnote
 {Title has been changed (previous title: Differential Geometry and
  Integrability of the Hamiltonian System of Closed Vortex Filament).}
 \bigskip\medskip\\
 \large    Norihito Sasaki\footnote
 {The present e-mail address {\tt h1899@ocugw.cc.osaka-cu.ac.jp},
  from which the present paper is sent to hep-th, is to be abrogated.
  New address is currently in petition (perhaps
  {\tt norih@sci.osaka-cu.ac.jp} will be assigned).}
 \smallskip\\
 \normalsize
 \it       Department of Physics, Osaka City University,
           Sumiyoshi-ku, Osaka 558, Japan\medskip\\
 \rm       October 1995\quad (revised, March 1996)
\end{center}
\bigskip\bigskip
 
\noindent
{\bf Abstract.}
The system of a closed vortex filament is an integrable Hamiltonian one,
namely, a Hamiltonian system with an infinite sequense of constants of
motion in involution. An algebraic framework is given for the aim of
describing differential geometry of this system. A geometrical structure
related to the integrability of this system is revealed. It is not a
bi-Hamiltonian structure but similar one. As a related topic, a remark
on the inspection of J.~Langer and R.~Perline, J.~Nonlinear Sci.~1, 71
(1991), is given.
\bigskip\bigskip
 
\bigskip\bigskip
 
\section{Introduction}
\label{sect:intro}
The vortex filament equation, which arises as the equation of motion
for the vorticity in a certain approximation of 3-dimensional fluid
forming a filament \cite{R:H}, has a description as an infinite
dimensional nonlinear Hamiltonian system \cite{M-W}. This is an
integrable system in the sense that there exists an infinite sequense
of constants of motion in involution, namely, Poisson-commutative
functions including the Hamiltonian function \cite{L-P}. To assert
this statement rigorously, some boundary condition or restriction
on the configuration of filament should be imposed relevantly.
 
As a relevant phase space of the filament or Poisson manifold
for this system, J.~Langer and R.~Perline \cite{L-P} introduced $\BAL$,
which consists of the points represented by a single curve (position
of the vortex filament) with infinite length, non-vanishing curvature
and asymptotic convergence to a fixed line.
They studied the Poisson structure of $\BAL$ and defined the sequense
of functions (or functionals) $I_{-2}$, $I_{-1}$, $\ldots$ in involution
by means of relating, via the Hasimoto map \cite{Has}, the system of
vortex filament to that of the nonlinear Schr\"odinger (NLS) equation
\cite{F-T}, a well-known and well-investigated integrable Hamiltonian
system.
The Hamiltonian system of NLS equation has a bi-Hamiltonian structure,
which is known as a geometrical background supporting the integrability
\cite{G-Do,Mag}. Although it is not quite obvious, the connection
of the filament of $\BAL$ to the NLS seems to suggest that $\BAL$
also has a bi-Hamiltonian structure.
 
Let $\kappa$, $\tau$ and $s$ denote the curvature, torsion and
arclength, respectively, along the curve, {\it cf.}~eq.~(\ref{eq:FSrel}).
Let $\kappa^{(i)}:=(\Ders)^i\kappa$ and $\tau^{(i)}:=(\Ders)^i\tau$.
We refer to a polynomial in $\kappa^{(i)}$ and $\tau^{(i)}$,
$i=0$, $1$, $\ldots$, with
coefficients in $\real$ as a local polynomial. We note that
the functions $I_{n-2}$ were expressed as $I_{n-2}=\int P_n \extd s$
with the local polynomials $P_n$,
\begin{equation}
 P_0 = 1, \quad
 P_1 = -\tau, \quad
 P_2 = \frac{1}{2}\, \kappa^2, \quad
 P_3 = \frac{1}{2}\, \kappa^2\tau, \quad
 P_4 = \frac{1}{2}\, \kappa^{(1)\,2}
      + \frac{1}{2}\,\kappa^2\tau^2 - \frac{1}{8}\,\kappa^4,
 \quad \ldots,
  \label{eq:polynomials}
\end{equation}
see \cite{L-P} for the definition.
 
In this paper we study the vortex filament with another restriction,
that is, to form a closed curve. As a phase space or
Poisson manifold for this system, we will introduce $\ECL$ (space
of Entirely Curved Loop).
 
The analysis parallel to that of J.~Langer and R.~Perline does not
work for $\ECL$, because in this case the image (in the space of
solutions to the NLS equation) of the Hasimoto map spans
a space in which the NLS system has not been well-investigated,
that is, the space of pseudoperiodic $\cplx$-valued functions with
various (not fixed) periods and phase-displacements around a period.
 
One can, however, observe the existence of infinite sequense of
functions on $\ECL$ in involution. Since every equation for local
polynomials makes sense locally on the curve without reflection of
the boundary condition, the easiest way to see this is to translate
results on $\BAL$ regarding local polynomials into statements on
$\ECL$, {\it cf.}~\Sect~\ref{sect:LPsF}.
It certainly explains the integrability of $\ECL$; nevertheless the
consideration within local polynomials seems awkward to clarify the
structure lying behind the integrability. Indeed, for both systems
$\BAL$ and $\ECL$, a recursive formula associated with the sequense
of functions in involution exists and is written in a simple form by
adopting indefinite integration, whose action cannot close within
the local polynomials.
 
Our investigation about $\ECL$ in order to elucidate a geometrical
structure related to the integrability is therefore adopting indefinite
integration and is made entirely within $\ECL$ itself.
\Sect~\ref{sect:LPsF} is exceptional; a remark related to $\BAL$ is
given there.
Although the space of dimension~3, in which the vortex filament live,
is originally Euclidean, remarkable properties hold for the vortex
filament in the space of constant curvature \cite{Koi,Y-O}. We follow
this generalization.
 
The paper is organized as follows: \Sect~\ref{sect:apf} is devoted to
giving an algebraic framework, which describes periodic functions
in an abstract manner in connection with integro-differential
operators. With the aid of this framework, we introduce
in \Sect~\ref{sect:ECL} algebraic objects that allow one
to define the differential calculus on $\ECL$.
A geometrical structure of $\ECL$ related to the integrability is
elucidated in \Sect~\ref{sect:calc}. This structure is related also
to the inspection reported in \cite{L-P} regarding the constants of
motion for $\BAL$. This is argued in \Sect~\ref{sect:LPsF}.
 
\section{Abstract Periodic Functions}
\label{sect:apf}
We introduce with an axiomatic definition the notion of
$\pdmod$-algebra, an abstraction of the algebra of
commutative-algebra-valued periodic functions.
 
We say the set $(\ALGEB,\:\{\intdif{m}\}_{m\in\rint})$ of a unital,
commutative and associative $\real$-algebra $\ALGEB$ and a sequence
of $\real$-linear operators $\intdif{m}$: $\ALGEB\rightarrow\ALGEB$
is a $\pdmod$-algebra if
\begin{equation}
 \intdif{m}\intdif{n}f = \intdif{m+n}f, \quad
 \intdif{1}(fg) = (\intdif{1}f)g + f(\intdif{1}g), \quad
 \intdif{-1}\big( (f-\intdif{0}f)g \big) = (f-\intdif{0}f)(\intdif{-1}g)
\end{equation}
for every $f$, $g\in\ALGEB$. Below, we refer simply to such an
algebra $\ALGEB$ as a $\pdmod$-algebra. Definite integration in a
$\pdmod$-algebra is defined by $\int := \id-\intdif0$.
 
\begin{example}
\label{exmp-dZA}
The algebra of smooth functions on $S^1:=\real/\rint$ has a
$\pdmod$-algebra structure. Let $f$ be such a function, {\it i.e.},
$f(x+1) = f(x)$ $\forall x\in\real$. Then $\intdif1$ is realized by
$(\intdif{1}f)(x) = \extd f(x)/\extd x$. And $\intdif{-1}$ is realized
by $\intdif{-1}f = \widetilde{\partial}{}^{0}\widetilde{\partial}{}^{-1}
\widetilde{\partial}{}^{0}f$,
where $(\widetilde{\partial}{}^{0}f)(x) = f(x)-\int_0^1 f(y)\extd y$
and $(\widetilde{\partial}{}^{-1}f)(x) = \int_0^x f(y)\extd y$.
\end{example}
 
Suppose $\ALGEB$ is a $\pdmod$-algebra. Then
$\FUNCT:=\int(\ALGEB)$ is shown to be a $\pdmod$-subalgebra of $\ALGEB$.
The operators $\intdif{m}$ act in $\FUNCT$ in the trivial way;
$\intdif{m}f = 0$, $\intdif{m}(fg) = f(\intdif{m}g)$ $\forall m\in\rint$,
$\forall f\in\FUNCT$, $\forall g\in\ALGEB$.
The formula $\int((\intdif{\pm 1}f)g + f(\intdif{\pm 1}g)) = 0$
$\forall f,\:g\in\ALGEB$ is often useful.
Note that the map $\int$: $\ALGEB\rightarrow\FUNCT$ is an (end)morphism
of $\real$-vector spaces, and not of algebras.
 
We say $\IDEAL$ is an ideal of $\pdmod$-algebra $\ALGEB$ if $\IDEAL$
is a not-unital $\pdmod$-subalgebra of $\pdmod$-algebra $\ALGEB$ and
is an ideal (as algebra) of algebra $\ALGEB$. Suppose this
situation and let $\FUNCT := \int(\ALGEB)$. Then, $\int(\IDEAL)$
coincides with $\IDEAL\cap\FUNCT$ and is an ideal of $\pdmod$-algebra
$\FUNCT$. The quotient algebra $\FUNCT/(\IDEAL\cap\FUNCT)$ can be
considered a subalgebra of $\ALGEB/\IDEAL$ and is identical to
the algebra $\int(\ALGEB/\IDEAL)$.
Note that the quotient space $\ALGEB/\IDEAL$ admits a natural
$\pdmod$-algebra structure induced from $\ALGEB$, so that
$\intdif{m}$ and $\int$ make sense as operators on $\ALGEB/\IDEAL$.
\begin{example}
\label{xmp:tracenull}
Let $\ALGEB$ be a $\pdmod$-algebra. Then, the subset
$\IDEAL := \{f\in\ALGEB \:|\: \int fg = 0$ $\forall g\in\ALGEB \}$
of $\ALGEB$ is an ideal of $\pdmod$-algebra $\ALGEB$.
\end{example}
 
\section{Formal Closed Curves}
\label{sect:ECL}
What we wish to describe is an unparametrized closed curve, but we start
with considering parametrized one, which is more convenient for the
variational calculus. The space where the curve lives is supposed to be
a 3-dimensional orientable Riemannian manifold $(\mfdM,\:\Riem{\:}{})$
of constant curvature; the curvature of $\mfdM$ is denoted by
$\curvM\in\real$, which is a given constant for the theory below.
The reader who is interested only in the case that $\mfdM$ is Euclidean
can simply set $\curvM = 0$ throughout the paper.
 
The curve can almost be specified by the curvature $\kappa$ and torsion
$\tau$ of the curve and the velocity $\lambda$ associated with the
parametrization. We will develop the theory based on these quantities.
This means 6 degrees of freedom, {\it i.e.}, gross position and direction
of the
curve, are neglected at the starting point, but this seems relevant for
discussing the integrability of the system of a single vortex filament.
We suppose $\lambda$ and $\kappa$ are non-vanishing at every point on
the curve.
 
We mention that we treat $\lambda$, $\kappa$ and $\tau$ formally
with forgetting their character of maps associating
a function on $S^1$ with a curve as I.~M.~Gel'fand and L.~A.~Dikii
have done in their formal variational calculus \cite{G-Do,G-Di};
$\lambda$, $\kappa$, $\tau$ and their derivatives are read
simply as indeterminates.  There are differences in several points
between our calculus and that of \cite{G-Do,G-Di}.
Among them, here we stress that our calculus is based
on the framework of $\pdmod$-algebra so that the formal curve is
implicitly of a closed shape.
 
Let $\algea$ be the unital, commutative and associative $\real$-algebra
generated by the indeterminates $\lambda^{(i)}$, $\kappa^{(i)}$,
$\tau^{(i)}$ ($i=0,\:1,\:\ldots$), $\lambda^{-1}$ and $\kappa^{-1}$
satisfying $\lambda^{-1}\lambda^{(0)}=\kappa^{-1}\kappa^{(0)}=1$.
We denote $\lambda := \lambda^{(0)}$, $\kappa := \kappa^{(0)}$ and
$\tau := \tau^{(0)}$.
The $\real$-linear map $\Ders$: $\algea\rightarrow\algea$ is defined by
$\Ders\lambda^{(i)}=\lambda^{(i+1)}$, $\Ders\kappa^{(i)}=\kappa^{(i+1)}$,
$\Ders\tau^{(i)}=\tau^{(i+1)}$ and Leibnitz rule
$\Ders(fg) = (\Ders f)g + f(\Ders g)$ $\forall f$, $g\in\algea$.
 
Identifying $\intdif1 = \lambda\Ders$, we introduce $\algeb$ as the
minimal extension of $\algea$ such that $\algeb$ is a $\pdmod$-algebra.
Let $\funct := \int(\algeb)$ and $\vectf := \algeb\laT\oplus
\algeb\laN\oplus\algeb\laB$. Here, the set of $\laT$, $\laN$ and $\laB$
stands for a certain basis of $\vectf$ orthonormal with respect to the
metric $\Riem{\:}{}$: $\vectf\times\vectf\rightarrow\algeb$.
The metric $\Riem{\:}{}$ is supposed $\algeb$-linear in each argument.
 
Let us define $\real$-linear operator $\nablas$: $\vectf\rightarrow\vectf$
by Leibnitz rule and Frenet-Serret relation
\begin{eqnarray}
&
 \nablas(f\laX) = ({\textstyle\Ders}f)\laX + f(\nablas\laX)\quad
 \forall f\in\algeb,\;\forall \laX\in\vectf,
 \label{eq:aaa}
& \\
&
 \nablas \laT = \kappa \laN, \quad
 \nablas \laN = - \kappa \laT + \tau \laB, \quad
 \nablas \laB = - \tau \laN.
 \label{eq:FSrel}
&
\end{eqnarray}
With this definition, one will identify $\laT$, $\laN$, $\laB$ as the
Frenet frame along the curve. The operator $\nablas$ is $\funct$-linear
and satisfies
\begin{equation}
 {\textstyle\Ders}\Riem{\laX}{\laY}
  = \Riem{\nablas\laX}{\laY} + \Riem{\laX}{\nablas\laY} \quad
 \forall \laX,\:\laY\in\vectf.
\end{equation}
 
Let us then introduce two operators $\delta_\laX$ and $\nablaof{\laX}$,
which are determined by the usual variational calculus as in \cite{L-P}
with regarding $\laX\in\vectf$ as a variational vector field.
For $\forall\laX\in\vectf$, the $\real$-linear operator $\delta_\laX$:
$\algeb\rightarrow\algeb$ is defined by
\begin{eqnarray}
&
 \delta_\laX(fg) = (\delta_\laX f)g + f(\delta_\laX g) \quad
  \forall f,\:g\in\algeb,
& \\
& \delta_\laX \intdif{m} = \intdif{m} \delta_\laX, \quad
 \delta_\laX {\textstyle\Ders} = {\textstyle\Ders}\delta_\laX
  - (\lambda^{-1}\delta_\laX\lambda){\textstyle\Ders},
& \\
& \delta_\laX\lambda = \Riem{\lambda \laT}{\nablas \laX},
& \\
& \delta_\laX\kappa = \Riem{\laN}{\nablas\nablas \laX + \curvM \laX}
 - \Riem{2\kappa \laT}{\nablas \laX},
& \\
& \delta_\laX\tau = {\textstyle\Ders}
    \Riem{\kappa^{-1}\laB}{\nablas\nablas \laX + \curvM \laX}
  + \Riem{\kappa\laB}{\nablas \laX} - \Riem{\tau\laT}{\nablas \laX}.
&
\end{eqnarray}
And, for $\forall\laX\in\vectf$, the $\real$-linear operator
$\nablaof{\laX}$: $\vectf\rightarrow\vectf$ is defined by
\begin{eqnarray}
&
 \nablaof{\laX}(f\laY) = (\delta_\laX f)\laY + f(\nablaof{\laX}\laY)\quad
  \forall f\in\algeb,\;\forall\laY\in\vectf,
& \\
& \nablaof{\laX}\laT
 = \Riem{\laN}{\nablas \laX}\laN + \Riem{\laB}{\nablas \laX}\laB,
& \\
& \nablaof{\laX}\laN
 = \Riem{\kappa^{-1}\laB}{\nablas\nablas \laX +\curvM \laX}\laB
 - \Riem{\laN}{\nablas \laX}\laT,
& \\
& \nablaof{\laX}\laB = - \Riem{\laB}{\nablas \laX}\laT
 - \Riem{\kappa^{-1}\laB}{\nablas\nablas \laX + \curvM \laX}\laN.
&
\end{eqnarray}
These operators satisfy
\begin{eqnarray}
&
 \delta_{\Phi \laX} = \Phi \delta_\laX,
 \quad \nablaof{\Phi \laX} = \Phi \nablaof{\laX} \quad
  \forall \Phi\in\funct,\; \forall\laX\in\vectf,
& \\
&
 \delta_\laX\Riem{\laY_1}{\laY_2}
  = \Riem{\nablaof{\laX}\laY_1}{\laY_2}
   + \Riem{\laY_1}{\nablaof{\laX}\laY_2} \quad
  \forall \laX,\:\laY_1,\:\laY_2\in\vectf.
 \label{eq:eee}
&
\end{eqnarray}
\begin{theorem}
The vector space $\vectf$ admits a Lie algebra structure
\begin{equation}
 [\laX,\:\laY] = \nablaof{\laX}\laY - \nablaof{\laY}\laX \quad
  \forall \laX,\:\laY\in\vectf.
 \label{eq:commutator}
\end{equation}
The algebra $\algeb$ is an $\vectf$-module with the action $\delta_\laX$,
$\laX\in\vectf$, namely,
\begin{equation}
 (\delta_\laX \delta_\laY - \delta_\laY \delta_\laX
  - \delta_{[\laX,\,\laY]})f = 0 \quad
   \forall \laX,\:\laY\in\vectf, \;\forall f\in\algeb.
 \label{eq:modulealg}
\end{equation}
\end{theorem}
This is proved through the following step: First, show the equation
(\ref{eq:modulealg}) for $f=\lambda$, $\kappa$, $\tau$ and then
extend (\ref{eq:modulealg}) to the whole $\algeb$. Second, verify the
Jacobi identity in $\vectf$ with the help of (\ref{eq:modulealg}).
Actual task is somewhat tedious but straightforward.
It is apparent that the algebra $\funct$ is an $\vectf$-submodule of
$\algeb$, since $\int$ commutes with $\delta_{\laX}$.
 
As was mentioned, the objects given above are related to a parametrized
curve. Now we will fix most of the ambiguity of parametrization by
imposing the relation $\Ders\lambda=0$, which means that the parameter
coincides with a multiple of the arclength. As a consequence, the
velocity will be eliminated except for its zero-frequency component
$\length := \int\lambda$, the circumferential length along the closed
curve.
 
Let $\ideal$ be the minimal ideal of $\pdmod$-algebra $\algeb$
such that $\Ders\lambda\in\ideal$. Let $\alge := \algeb/\ideal$,
$\func := \int(\alge)$, $\vect := \alge\liN\oplus\alge\liB$,
$\idealvf := \ideal\laT\oplus\ideal\laN\oplus\ideal\laB$, and
$\vectf/\idealvf := \alge\laT\oplus\alge\laN\oplus\alge\laB$.
 
There is no actual difference between the length $\length$ and the
velocity $\lambda$ in $\func$ or in $\alge$, because $\Ders\lambda = 0$.
However, we use both symbols; we prefer denoting it by $\length$ if and
only if we consider it as an element of $\func$ rather than of $\alge$.
 
Let $\canhom$ be the surjection $\vectf\rightarrow\vectf/\algeb\laT$
or surjection $\vectf/\idealvf\rightarrow\vect$,
\begin{equation}
 \canhom(f\laT + g\laN + h\laB) := g\liN + h\liB{}
\end{equation}
and let $\embed$ be the
injection $\funct\times(\vectf/\algeb\laT)\rightarrow\vectf$
or injection $\func\times\vect\rightarrow\vectf/\idealvf$,
\begin{equation}
 \embed(\Phi,\:g\liN + h\liB) :=
  (\intdif{-1}(\lambda\kappa g) + \Phi)\laT + g\laN + h\laB.
   \label{def-embed}
\end{equation}
The image $\vectf{}':=\embed(\funct,\:\vectf/\algeb\laT)$ of $\embed$
preserves $\ideal$, {\it i.e.},
$\delta_{\liX'}\ideal\subset\ideal$ $\forall\liX'\in\vectf{}'$, and is a
Lie subalgebra of $\vectf$. Hence $\alge$ is an $\vectf{}'$-module, so
is $\func$.
 
Recalling the definition of $\nablas$, we see that $\nablas$ preserves
$\idealvf$, {\it i.e.}, $\nablas(\idealvf)\subset\idealvf$, hence makes
sense as a map $\vectf/\idealvf\rightarrow\vectf/\idealvf$. Similarly,
we see that $\delta_{\liX'}$ and $\nablaof{\liX'}$ make sense as,
respectively, a map $\alge\rightarrow\alge$ and a map
$\vectf/\idealvf\rightarrow\vectf/\idealvf$ if $\liX'$ belongs to
$\vect{}':=\embed(\func,\:\vect)$.
Formulae (\ref{eq:aaa})--(\ref{eq:eee}) hold also for these quotient
spaces with the restriction $\liX'\in\vect{}'$ on $\liX'$ of
$\delta_{\liX'}$ and $\nablaof{\liX'}$.
 
As a corresponding result for these quotient spaces, the image $\vect{}'$
of $\embed$ in $\vectf/\idealvf$ is a quotient Lie algebra of $\vectf{}'$,
$\pdmod$-algebra $\alge$ is an $\vect{}'$-module and the
subalgebra $\func$ is an $\vect{}'$-submodule.
 
Furthermore, it is not difficult to show that $\func\laT$ is an ideal of
Lie algebra $\vect{}'$ and that $\delta_{\laT}f = \Ders f$
$\forall f\in\alge$, thus,
\begin{proposition}
The vector space $\vect$ admits a Lie algebra structure
\begin{equation}
 [\liX,\:\liY] := \canhom(\nablaof{\liX'}\liY' - \nablaof{\liY'}\liX'),
 \quad \liX' := \embed(0,\:\liX),\quad \liY' := \embed(0,\:\liY)
 \quad \forall \liX,\:\liY\in\vect.
\end{equation}
The algebra $\func$ is an $\vect$-module with the action
$\delta_{\embed(0,\:\liX)}$ of $\liX\in\vect$.
\end{proposition}
 
We call an element of $\func$ a function and an element of $\vect$
a vector field.
Following the standard notation of differential geometry, we denote
the action of vector field $\liX$ to the function $\Phi$ by
the left action $\liX\Phi := \delta_{\embed(0,\:\liX)}\Phi$.
Since $\vect$ is an $\alge$-vector space, function
$\Phi\in\func\subset\alge$ can naturally act to the vector field $\liX$;
we denote it by the left action $\Phi\liX$.
 
As we have defined or shown, $\func$ is a commutative and associative
algebra, $\vect$ is a Lie algebra, $\func$ is a left $\vect$-module and
$\vect$ is a left $\func$-module. It is obvious that each vector field
$\liX\in\vect$ acts as a derivation to the product either $\Phi\Psi$
or $\Phi\liY$ $\forall\Phi$, $\Psi\in\func$, $\forall\liY\in\vect$.
In addition, $(\Phi\liX)\Psi = \Phi(\liX\Psi)$ holds for $\forall\Phi$,
$\Psi\in\func$, $\forall\liX\in\vect$.
These are condition enough to introduce, as in \Sect~\ref{sect:calc},
the differential calculus.
 
Let us introduce a Riemannian metric for $\vect$. Although we have no
proof, it seems true that
\begin{equation}
  \int\! fg = 0\;\forall g\in\alge\;\;\Rightarrow\;\; f = 0
  \label{eq:nondeg}
\end{equation}
for $\forall f\in\alge$. Absense of the proof is not so serious, because
it is at least possible to impose this rule; if (\ref{eq:nondeg}) is
false, we can redefine $\alge$ (and accordingly $\func$, $\vect$) to
turn (\ref{eq:nondeg}) into true with preserving the $\pdmod$-algebra
structure ({\it cf.}~Example~\ref{xmp:tracenull}) and,
in addition, operators introduced in this section make sense for
these redefined objects. Therefore, in particular, the inner product
$(\:,\:)$: $\vect\times\vect\rightarrow\func$,
\begin{equation}
 (g_1 \liN + h_1 \liB,\: g_2 \liN + h_2 \liB) = \int \lambda(g_1 g_2
  + h_1 h_2)  \quad \forall g_1,\:g_2,\:h_1,\:h_2\in\alge
   \label{eq:inner}
\end{equation}
is supposed non-degenerate and is thought of as a Riemannian structure.
Then, for given any function $\Phi$, there uniquely exists
a vector field $\liX$ (the gradient of $\Phi$) such that
$(\liX,\:\liY) = \liY\Phi$ $\forall\liY\in\vect$.
 
The parametrization of the curve have been fixed except for the ambiguity of
rigid slide by imposing the relation $\Ders\lambda = 0$. And there are no
elements in $\func$ (and even in $\funct$) that can distinguish
the difference within this ambiguity. The objects $\func$ and $\vect$
can, hence, be thought of as the algebra of functions and Lie algebra
of vector fields, respectively, on (a certain quotient space of)
the space consisting of the points represented by an
unparametrized closed curve with non-vanishing length $\length$ and
non-vanishing curvature $\kappa$ living in a 3-dimensional
space $\mfdM$ of constant curvature equal to $\curvM$. We denote the set
of these algebraic objects by $\ECL(\curvM)$ or simply by $\ECL$. We use
phrases as if $\ECL$ denoted a Riemannian manifold.
 
\section{A Structure Related to the Integrability}
\label{sect:calc}
We denote by $\otimes$ the operation making $\func$-tensor product.
Let $\forms{p}$ denote the vector space of $\func$-linear maps
$\omega$: $\vect^{\otimes p}\rightarrow\func$ such that
$f := \omega(\liX_1\otimes\cdots\otimes\liX_p)$, $\liX_i\in\vect$,
is skew-symmetric, if $p\geq2$, under
the exchange of $\liX_i$ and $\liX_j$, $i\neq j$, and has algebraic
expression $f = \int(\cdots)$ written with the ingrediens $g_i$, $h_i$
($g_i\liN + h_i\liB := \liX_i$), $\omega$-dependent elements of $\alge$
and operators $\intdif{m}$.
Such an element $\omega$ of $\forms{p}$ is called a $p$-form.
 
The exterior derivative $\extd$: $\forms{p}\rightarrow\forms{p+1}$
and the exterior product
 $\wedge$: $\forms{p}\times\forms{q}\rightarrow\forms{p+q}$
\linebreak are introduced algebraically in the usual manner.
To fix the convention, we present \linebreak three formulae \hfill
$(\extd\Phi)(\liX) = \liX\Phi$, \hfill $(\extd\omega)(\liX\otimes\liY)
 = \liX(\omega(\liY))-\liY(\omega(\liX)) - \omega([\liX,\:\liY])$
\hfill and \linebreak $(\omega \wedge \eta)(\liX \otimes \liY)
 = \omega(\liX)\eta(\liY) - \omega(\liY)\eta(\liX)$
$\forall \liX,\:\liY\in\vect$, $\forall\Phi\in\forms{0}=\func$,
$\forall\omega,\:\eta\in\forms{1}$.
 
Given 1-form $\xi\in\forms{1}$, we associate a vector field
$\dual\xi\in\vect$ by $(\dual\xi,\:\liY) = \xi(\liY)$
$\forall\liY\in\vect$ with the Riemannian structure (\ref{eq:inner}).
Given operator $H$: $\vect\rightarrow\vect$, we
associate an operator $H\dual$: $\forms{1}\rightarrow\vect$
by $(H\dual) \xi = H(\dual\xi)$ $\forall\xi\in\forms{1}$.
Operator $H$: $\vect\rightarrow\vect$ is called a skew-adjoint
operator if $H$ is $\func$-linear and obeys
$(\liX,\:H\liY)=-(H\liX,\:\liY)$ $\forall\liX,\:\liY\in\vect$.
 
Here, we briefly review geometrical notions and statements regarding
(bi-)Hamiltonian structure \cite{G-Do}, see also \cite{Mag}.
Given two skew-adjoint operators $H_1$ and $H_2$, define the 3-form
$[H_1,\:H_2]\in\forms{3}$,
\begin{eqnarray}
&& [H_1,\:H_2](\liX\otimes\liY\otimes\liZ)
 \nonumber\\
&=&
 \Bigl\{ (H_1\liX)(H_2\liY,\:\liZ) + (\liZ,\:[H_1\liY,\:H_2\liX])
  + (H_1\leftrightarrow H_2) \Bigr\} + \cycle(\liX,\:\liY,\:\liZ)
 \label{eq:defSc}\\
&& \qquad\qquad\qquad\qquad\qquad\qquad
   \qquad\qquad\qquad\qquad\qquad\qquad
 \forall \liX,\:\liY,\:\liZ\in\vect. \nonumber
\end{eqnarray}
The 3-form $[H_1,\:H_2]$ is the dual (for notational simplicity) of
the Schouten bracket between $H_1\dual$ and $H_2\dual$.
A skew-adjoint operator $J$ (or $J\dual$ \cite{G-Do}) is called a
Hamiltonian operator if $[J,\:J]=0$. A Hamiltonian operator $J$ induces
a Poisson structure
\begin{equation}
 \{\Phi,\:\Psi\} = (J\dual \extd\Phi)\Psi \quad
  \forall \Phi,\:\Psi\in\func
  \label{eq:PB}
\end{equation}
and $J\dual\extd$ provides a morphism of Lie algebras
$\func\rightarrow\vect$.
A pair of two Hamiltonian operators $J$ and $H$ such that $[J,\:H]=0$
is called a bi-Hamiltonian structure (or Hamiltonian pair \cite{G-Do}),
which implys, with certain further conditions, the existence of a
sequence of functions in involution or Poisson-commutative functions.
 
Let us return to the problem of $\ECL$.
Define operator $J$: $\vect\rightarrow\vect$,
\begin{equation}
 J(g \liN + h \liB) = h \liN - g \liB \quad \forall h,\:g\in\alge
  \label{eq:defJ}
\end{equation}
and operator $K$: $\vect\rightarrow\vect$,
\begin{equation}
 K(\liX) = J\,\canhom\,\nablas\,\embed(0,\:J\liX)
  \quad\forall\liX\in\vect.
  \label{eq:defK}
\end{equation}
It is easy to show that the operators $J$ and $K$ are skew-adjoint.
These are operators for $\ECL$ analogous to those for $\BAL$
given in \cite{L-P}, where J.~Langer and R.~Perline showed that
the operators corresponding to $J$ and $K$ above generate recursively
the sequense of vector fields on $\BAL$ associated with
certain functions in involution. Below, we will consider a recursively
generated sequense of 1-forms on $\ECL$ defined analogously.
 
Let $\xi_n$, $n=0$, $1$, $\ldots$ be the sequense of 1-forms defined by
\begin{equation}
 \xi_n := \length^{n-1}\extd\length\circ(K J^{-1})^n,
 \quad \mbox{\it i.e.},\;
\dual\xi_n := \length^{n-1}(J^{-1}K)^n (-\kappa\liN).
 \label{eq:defxis}
\end{equation}
\begin{theorem}
\label{th:SchBr}
The dual (\ref{eq:defSc}) of the Schouten brackets for skew-adjoint
operators $J$ and $K$ defined by (\ref{eq:defJ}) and (\ref{eq:defK}),
respectively, are
\begin{eqnarray}
[J,\,J](\liX\otimes \liY\otimes \liZ) &=& 0, \label{eq:Sc-JJ}\\ \relax
[J,\,K](\liX\otimes \liY\otimes \liZ) &=& 2\xi_0(J \liX)\,(\liY,\: K \liZ)
  + \cycle(\liX,\:\liY,\:\liZ), \\ \relax
[K,\,K](\liX\otimes \liY\otimes \liZ) &=&
 2\xi_0(J \liX)\,(\liY,\:(3 K J^{-1} K - \curvM J)\liZ)
  + \cycle(\liX,\:\liY,\:\liZ) \label{eq:Sc-KK}
\end{eqnarray}
with $\xi_0$ of (\ref{eq:defxis}).
\end{theorem}
The proof we know is not so easy nor so straightforward. However, we
would like to omit presenting it, since it requires a tedious
calculation and no tricky technics.
 
The theorem above says $J$ is a Hamiltonian operator but the pair $J$
and $K$ is not of a bi-Hamiltonian. It is implied by investigations
of \cite{M-W,Bry} that $J$ is a Hamiltonian operator.
The vortex filament equation is expressed by the Hamiltonian function
$\length$ with the Poisson bracket induced from $J$ \cite{M-W,Bry}.
 
\begin{theorem}
\label{th:deronef}
The sequense of 1-forms $\xi_n$ defined by (\ref{eq:defxis}) satisfies
\begin{equation}
 \extd\xi_n = \sum_{k\in\rint,\;0 < 2k < n} (2k - n)(\xi_{k}\wedge
  \xi_{n-k} - \cMll \xi_{k-1}\wedge\xi_{n-k-1}).
 \label{eq:strictrel}
\end{equation}
\end{theorem}
Before sketching the proof, we note that the equation
\begin{eqnarray}
&& 2\,\Bigl( \extd\xi \circ(H_1\otimes H_1)
 - \extd\eta \circ(H_1\otimes H_2 + H_2\otimes H_1)
 + \extd\zeta \circ(H_2\otimes H_2) \Bigr) (\liX\otimes\liY)
 \nonumber \\
&=&  [H_1,\:H_1](\dual\xi \otimes\liX \otimes\liY)
  - 2\, [H_1,\:H_2](\dual\eta \otimes\liX \otimes\liY)
  + [H_2,\:H_2](\dual\zeta \otimes\liX \otimes\liY)
 \label{eq:useSc} \\
&& \qquad\qquad\qquad\qquad\qquad\qquad\qquad
   \qquad\qquad\qquad\qquad\qquad\qquad\qquad
   \forall \liX,\:\liY\in\vect \nonumber
\end{eqnarray}
is fulfilled for any skew-adjoint operators $H_1$ and $H_2$ and
1-forms $\xi$, $\eta$, $\zeta$ whenever
\begin{equation}
 H_1\dual\xi = H_2\dual\eta, \quad H_1\dual\eta = H_2\dual\zeta.
\end{equation}
Taking $H_1=J$, $H_2=K$, $\xi = \length^{2-n}\xi_{n+2}$,
$\eta = \length^{1-n}\xi_{n+1}$ and $\zeta = \length^{-n}\xi_{n}$ in
(\ref{eq:useSc}) and using Theorem~\ref{th:SchBr}, we can derive a
recursive relation for $\extd\xi_k$.
The recursive relation is simplified by
\begin{equation}
 \xi_i(J\dual\xi_j) = 0, \quad \xi_i(K\dual\xi_j) = 0,
  \label{eq:cons-skew}
\end{equation}
which follow from skew-adjointness of $J$ and $K$. Using this recursive
relation, one can prove Theorem~\ref{th:deronef} by induction with the
initial data $\extd\xi_0 = \extd\xi_1 = 0$, which are easily seen from
$\xi_0 = \length^{-1}\extd\length$ and $\xi_1 = \extd\int(-\lambda\tau)$.
 
As we will see below, the relation (\ref{eq:strictrel}) is to strongly
suggest the integrability.
\begin{proposition}
Suppose the triviality of the de~Rham cohomology of degree 1, {\it i.e.},
suppose every closed 1-form is exact. Let a function
$\cMll$ and 1-forms $\xi_0$, $\ldots$, $\xi_m$
be given. If these satisfy (\ref{eq:strictrel}) and
$\extd(\cMll)=2\cMll\xi_0$,
then there exist functions $f_1$, $\ldots$, $f_m$ such that
\begin{equation}
 \extd f_n = \xi_n + \sum_{k\in\rint,\;1<k<n} (1-k)(f_k \xi_{n-k}
  - \cMll f_{k-1}\xi_{n-k-1}), \quad 1\leq n\leq m.
  \label{eq:oneans}
\end{equation}
\end{proposition}
This is easily proved by induction in $m$. The proposition is
of general character; it is not necessary to suppose that the
differential calculus is that on $\ECL$.
 
If there exist functions $f_1$, $f_2$, $\ldots\in\func$ satisfying
(\ref{eq:oneans}) for $\xi_n$ of (\ref{eq:defxis}), it is readily
seen from (\ref{eq:cons-skew}) and $\func$-linearity of $J$ that
these functions and $\length$ are in involution. Moreover, it is
not difficult to show that $\xi_0$, $\xi_1$, $\ldots$ are linearly
independent over $\func$, so are the flows associated, via
$J\dual\extd$, with the functions $\length$, $f_1$, $f_2$, $\ldots$
(if exist).
 
We do not know whether the de~Rham cohomology of degree~1 for $\ECL$
is trivial or not. Therefore, the existence of these functions on
$\ECL$ is, rigorously speaking, not proved. However, their existence
itself seems true; we conjecture that
\begin{equation}
 f_n = \length^{n-1}\sum_{k\in\rint,\;0\leq 2k\leq n}
  \frac{(2k)!}{2^{2k}(k!)^2} \, \curvM^k \int \lambda P_{n-2k}
\end{equation}
with (\ref{eq:polynomials}) solve the equation (\ref{eq:oneans}),
{\it cf.}~\Sect~\ref{sect:LPsF} for the case $\curvM = 0$.
 
We close this section with the following remarks. Recalling the argument
above, we pick out the set $(J,\:K,\:\length)$ of invertible Hamiltonian
operator $J$, skew-adjoint operator $K$ and non-vanishing function
$\length$ as a geometrical structure of $\ECL$. Theorem~\ref{th:SchBr}
and the equation $\extd(\extd\length\circ(K J^{-1})) = 0$ are, then,
essential for the integrability. This structure is enough to define
recursively a sequense of 1-forms, whose external derivatives
obey the relation (\ref{eq:strictrel}), and suggests the
existence of the sequense of functions in involution.
This seems to indicate a possibility of generalizing
bi-Hamiltonian structure into something relaxed but preserving the
nature supporting the integrability.
 
\section{A Remark on Langer-Parline's Formula}
\label{sect:LPsF}
Let us introduce a $\rint$-grading to $\algeb$, $\funct$, $\vectf$,
$\alge$, $\func$, $\vect$ and $\forms{p}$ as follows.
Let $\deg\real = 0$, $\deg\lambda = -1$, $\deg\kappa = \deg\tau = 1$,
$\deg\curvM = 2$ (with regarding $\curvM$ as an indeterminate),
$\deg \laT = \deg \laN = \deg \laB = 0$. Let operators/maps $\intdif{m}$,
$\int$ and $\canhom$ do not change the degree. This grading is compatible
with operators $\delta_{\laX}$, $\nablaof{\laX}$, {\it etc}. It is easily
shown that the external derivative $\extd$ increses the degree by 1.
This is used below.
 
As was mentioned in \Sect~\ref{sect:intro}, results on $\BAL$
regarding local polynomials can be translated into the statements
for $\ECL$. We will do so, but we would like to note that this relys
on the assumptions (i) the convergence of improper integrals posed
in \cite{L-P} and (ii) harmlessness of reading $\kappa^{(i)}$ and
$\tau^{(i)}$, which in \cite{L-P} are variables depending on the
curve of $\BAL$, as indeterminates.
 
Below we set $\curvM = 0$ for simplicity, but (\ref{eq:dlpsfm}) can be
derived even if $\curvM\neq 0$ with the help of a hint in \cite{Y-O}.
 
We can read (A)--(C) in \cite{L-P}:
Let $\liX_{n-2} := J\dual\extd \int\lambda P_n \in\vect$ with $P_n$ of
(\ref{eq:polynomials}), where (A) $P_n$ are local polynomials. (B) There
exist local polynomials $P_{\laT,\:n}$, $P_{\laN,\:n}$ and $P_{\laB,\:n}$
such that $\liX_{n-2} = P_{\laN,\:n}\liN + P_{\laB,\:n}\liB$ and
$\Ders P_{\laT,\:n} = \kappa P_{\laN,\:n}$. We see
$\deg P_n = \deg P_{\laT,\:n-1} = n$ with fixing the ambiguity of
$P_{\laT,\:n}$ appropreately. Then, (C)
\begin{equation}
 \liX_{(n+1)-2} = J\,\canhom\,\nablas (P_{\laT,\:n}\laT
    + P_{\laN,\:n}\laN + P_{\laB,\:n}\laB).
\end{equation}
Note that the right hand side slightly differes from $KJ^{-1}\liX_{n-2}$
with $J$ of (\ref{eq:defJ}) and $K$ of (\ref{eq:defK}).
 
{}From these translated facts, we see $KJ^{-1}\liX_{n-2} - \liX_{(n+1)-2}
  = (\int P_{\laT,\:n}) J(-\kappa \liN)$, which results in
\begin{equation}
 \zeta_n = \extd\int\lambda P_n
  + \length^{-1}\sum_{k\in\rint,\;2\leq k\leq n} (\int\lambda
   P_{\laT,\:k-1})\zeta_{n-k},  \label{eq:resin}
\end{equation}
where we have put $\zeta_n := \length^{1-n}\xi_n$. Using this relation
and (\ref{eq:strictrel}) with $\curvM = 0$, we can show that
\begin{equation}
 \sum_{k\in\rint,\;2\leq k\leq n} \extd\Delta_k\wedge\zeta_{n-k} = 0,
 \quad \Delta_k := \int\lambda((k-1)P_k - P_{\laT,\:k-1}).
\end{equation}
This equation for $n = 2$ implies that there exists $\Phi\in\func$
such that $\extd\Delta_2 = \Phi\zeta_0$. We note that there exist
local polynomials $P_\laN$ and $P_\laB$ and some integer $m$ such that
$\dual\extd\int\lambda P = \kappa^{-m}(P_\laN \liN + P_\laB \liB)$ for
every local polynomial $P$. Since $\dual\zeta_0 = -\kappa\liN$, we see
$\Phi\kappa^{m+1}$ for some integer $m$ is also a local polynomial,
while $\Phi\in\func$, hence $\Phi\in\real$. We now have two ways of
evaluating $\deg\Phi$, namely, $\deg\Phi = 1 + \deg\Delta_2
 - \deg\zeta_0 = 2$ and $\deg\Phi = \deg(\real) = 0$.
This mismatch is permissible only if $\Phi = 0$,
hence $\extd\Delta_2 = 0$. The argument can be extended iteratively
to assert $\extd\Delta_3 = \extd\Delta_4 = \cdots = 0$, {\it i.e.},
\begin{equation}
 (n-1) \int\lambda P_n \equiv \int\lambda P_{\laT,\:n-1},
 \quad n = 2,\:3,\ldots,
  \label{eq:dlpsfm}
\end{equation}
where $\,\equiv\,$ denotes the equivalence in the image in $\forms{1}$
of $\extd$. This proves significant part of the Langer-Perline's
inspection \cite{L-P}, which says the validity even in $\func$ of the
formula (\ref{eq:dlpsfm}).
 
\section*{Acknowledgements}
The author is grateful to Dr.~Y.~Yasui for profitable discussions.

\end{document}